\documentclass[useAMS,usenatbib]{mn2e}
\usepackage{amsmath}
\usepackage{amssymb}
\usepackage{graphicx}

\usepackage[pdftex]{hyperref}   
\hypersetup{bookmarks=true,colorlinks=true,linkcolor=black,citecolor=blue,urlcolor=blue}

\title[\emph{Gaia}'s potential for the discovery of circumbinary planets]{\emph{Gaia}'s potential for the discovery of circumbinary planets}
\author[Sahlmann, Triaud, \& Martin]
{J. Sahlmann$^{1}$\thanks{E-mail: Johannes.Sahlmann@esa.int}, 
A. H. M. J. Triaud$^{2}$\thanks{Fellow of the Swiss national science foundation}, 
D. V. Martin$^{3}$
\\
$^{1}$European Space Agency, European Space Astronomy Centre, P.O. Box 78, Villanueva de la Ca\~nada, 28691 Madrid, Spain\\
$^{2}$Kavli Institute for Astrophysics \& Space Research, Massachusetts Institute of Technology, Cambridge, MA 02139, USA\\
$^{3}$Observatoire de Gen\`eve, Universit\'e de Gen\`eve, 51 Chemin Des Maillettes, 1290 Versoix, Switzerland
}
\begin{document}

\date{Accepted 2014 mmm dd. Received 2014 mmm dd; in original form 2014 August 25}

\pagerange{\pageref{firstpage}--\pageref{lastpage}} \pubyear{2014}

\maketitle

\label{firstpage}

\begin{abstract}
The abundance and properties of planets orbiting binary stars -- circumbinary planets -- are largely unknown because they are difficult to detect with currently available techniques. Results from the \emph{Kepler} satellite and other studies indicate a {minimum} occurrence rate of circumbinary giant planets of $\sim$10 \%, yet only a handful are presently known. Here, we study the potential of ESA's \emph{Gaia} mission to discover and characterise extrasolar planets orbiting nearby binary stars by detecting the binary's periodic astrometric motion caused by the orbiting planet. We {expect} that \emph{Gaia} will discover hundreds of giant planets around binaries with FGK dwarf primaries within 200 pc of the Sun, if we assume that the giant planet mass distribution and abundance are similar around binaries and single stars. {If on the other hand all circumbinary gas giants have masses lower than two Jupiter masses, we expect only four detections. \emph{Gaia} is critically sensitive to the properties of giant circumbinary planets and will therefore make the detailed study of their population possible.} {\it Gaia}'s precision is such that the distribution in mutual inclination between the binary and planetary orbital planes will be obtained. It also possesses the capacity to establish the frequency of planets across the H-R diagram, both as a function of mass and of stellar evolutionary state from pre-main sequence to stellar remnants. {\it Gaia}'s discoveries can reveal whether a second epoch of planetary formation occurs after the red-giant phase.

\end{abstract} 


\section{Introduction}

A number of systems composed of planets orbiting both components of a binary - circumbinary planets - have been detected in recent years. This includes systems detected from radial velocity data such as HD\,202206 \citep{Correia:2005lr}, such as NN Serpentis using eclipse timing variations (e.g. \citealt{Beuermann:2010fk, Marsh:2013lr}), like Kepler-16 detected from transits by the {\it Kepler} spacecraft (e.g. \citealt{Doyle:2011vn}), like Ross 458, a planetary-mass object directly imaged around an M-dwarf binary \citep{Burgasser:2010aa}, {and like PSR B1620$-$26, a pulsar$+$white-dwarf binary that hosts a Jupiter-mass planet \citep{Thorsett:1999aa, Sigurdsson:2008aa}}.

While the reality of circumbinary planets was for a time questioned, the small harvest of systems identified by {\it Kepler} indicated these systems are fairly common. The occurrence rate of gas giants in {the} circumbinary configuration has been estimated to be of order 10\% \citep{Armstrong:2014aa,Martin:2014aa}, similar to the abundance of gas giants orbiting single stars \citep{Cumming:2008ys,Mayor:2011fj}. This came a bit as a surprise since planet formation mechanisms were thought to be stifled by mixing caused by the binary into the protoplanetary disc (e.g. \citealt{Meschiari:2012qy}), or by the very formation mechanism of close binaries \citep{Mazeh:1979eu,Fabrycky:2007pd}.

{\it Gaia} \citep{Perryman:2001lr} {was launched in December 2013 and began its survey in July 2014}. Over the next five years, it will survey the entire sky and measure the positions {and motions} of approximately one billion stars. Individual measurements will be collected with a precision of $\sim$30 {micro-arcseconds} ($\mu$as) {for sources} with apparent {optical} magnitudes brighter than 12. This is sufficient to allow for the detection of thousands of planets in nearby systems. {\cite{Casertano:2008th} and} \citet{Sozzetti:2014qy} investigated the potential for {\it Gaia} to detect gas giants using the astrometric method, {and} \cite{Neveu:2012fk} {studied} the interest of combining astrometric and radial-velocity {data}, very much like what {was} achieved for brown dwarfs using {\it Hipparcos} \citep{Sahlmann:2011qy}. However, this body of work only considered planets orbiting single stars.

It is reasonable to think that a large fraction of the systems where {\it Gaia} has the sensitivity to detect gas giants will be binary star systems, a number of which will have fairly short separations. In this paper, we look into whether gas giants in circumbinary orbits can be discovered using {\it Gaia}'s measurements. First we describe the importance of {\it Gaia}'s contribution to the study of circumbinary  planets. Then, in Sect.~\ref{sec:gaia}, we present the type, precision, and number of measurements we can hope to obtain by the end of {\it Gaia}'s nominal mission, as well as the orbital elements required to be calculated to make a detection. {We study some }benchmark examples, before expanding our study to the nearby population of binaries in Sect. \ref{sec:simulation}. Finally, we discuss our results.

\section{Science case}

Astrometric planet detection has limitations that are very different from the other four observing techniques currently employed in the discovery of circumbinary planets: {transit photometry, eclipse-timing or transit-timing variations, radial velocimetry, and direct imaging}. The most important distinction is the wide range of binary star properties that is accessible for planet detection. Transit photometry, eclipse-timing variations, and radial velocimetry are mostly confined to the shortest period binaries, whereas direct imaging can mostly access common proper motion companions on extremely wide orbits.  

Direct imaging observations are most sensitive to self-luminous planets that can be resolved next to the binary host and so far revealed planetary-mass objects at orbital separations of 80-1000 AU around two young and low-mass M-dwarf binaries \citep{Goldman:2010aa, Scholz:2010aa, Burgasser:2010aa, Delorme:2013aa}. 

Transit and eclipse probabilities have strongly decreasing functions with orbital period. Consequently, transiting planet discoveries have been limited to sub-AU separations around binaries with periods of a few tens of days at most. In addition, the region immediately outside a binary's orbit is unstable, which further restrains the available parameter space. Stable orbits start appearing for circumbinary periods greater than $\sim$4.5 times the binary period \citep{Dvorak:1986aa,Dvorak:1989vn,Holman:1999lr,Pilat-Lohinger:2003qy}.  
Radial-velocity surveys are also best suited to short-period systems and are impacted by the instability region (e.g. \citealt{Konacki:2009aa}). The method of eclipse-timing variations (ETV) is sensitive to long-period planets {($\gtrsim 200$ days) since the ETV amplitude increases with the planet period \citep{Borkovits:2011aa}. Whilst the ETV amplitude is independent of the binary period, short-period binaries ($\lesssim 50$ days) are favoured since they allow us to }reach the temporal precision required to detect the light travel time effect induced by a planet. 

Astrometry, therefore, is a complementary technique: its sensitivity improves with lengthening periods making it less affected by the instability region. This implies that a wide range of binary periods can be included in the search. Furthermore, unlike the other techniques, an astrometric survey can reach any binary (eclipsing or not), and allows the calculation of all binary and planetary orbital elements. Particularly, it is not impaired by a particular orbital plane orientation on the sky: true planetary masses can be measured and no transits are required for detection. Its limitations principally come from the distance to the system, the precision on individual measurements, and the timespan of the survey itself. 

Results from {\it Gaia} will help us understand better the systems discovered by the {\it Kepler} satellite, which has provided the largest and best-understood sample of circumbinary planets. A degeneracy exists between the abundance of circumbinary gas giants and the distribution in mutual inclination between the planetary and binary orbital planes. Only a minimum frequency { of $\sim10\%$} could be inferred from the {\it Kepler} systems \citep{Armstrong:2014aa,Martin:2014aa}, { corresponding to a coplanar distribution of planets. If  circumbinary planets instead exist on a range of mutual inclinations, this abundance increases, e.g. a Gaussian distribution with standard deviation of $5^{\circ}$ corresponds to an abundance of $20\%$. The latest observations suggest such a scenario: Kepler-413 has a mutual inclination of $4.1^{\circ}$ and two candidates found by the ETV method have mutual inclinations of $10$ and $19^{\circ}$ (Welsh et al. presented at Litoms\v yl, Czech Republic, 2014).}

{ The astrometric method can measure the distribution of mutual inclinations and hence produce a true occurrence rate}. Observational evidence points out that circumbinary discs can be partially inclined \citep{Winn:2006qy,Plavchan:2008fk,Plavchan:2013lr}, and that gas giants  in circumstellar orbits are regularly found on inclined orbits \citep{Schlaufman:2010fk,Winn:2010rr,Triaud:2010fr,Albrecht:2012lp}. The prevalence of inclined systems would be an important marker of the formation and dynamical evolution of circumbinary planets. 

Several massive and long period planets have been claimed from eclipse-timing variations orbiting post-common envelope binaries (see  \citealt{Beuermann:2012fk} for a list). It is unclear whether these planets existed during the main-sequence or whether they formed as a second generation using matter ejected from their hosts \citep{Schleicher:2014qy}. \citet{Mustill:2013lr} notably show that it is hard to recover NN Ser's architecture from a dynamical evolution of the system influenced by heavy mass loss. {\it Gaia} will survey binaries like NN Ser and systems that are similar to the progenitors of these binaries. {\it Gaia} will test results from eclipse-timing variations and permit to study the architecture and mass evolution of planetary systems pre and post red-giant phase. Thanks to its sensitivity, it may even be able to independently verify  the existence of some of the currently claimed systems (see Sect. \ref{sec:discussion}).

\section{\emph{Gaia} astrometric measurements}\label{sec:gaia}

The \emph{Gaia} satellite is performing an all-sky survey in astrometry, photometry, and spectroscopy of all star-like objects with \emph{Gaia} magnitudes $G\la20$ \citep{Perryman:2001lr, de-Bruijne:2012kx}, {where the $G$ passband covers the 400--1000 nm wavelength range with maximum transmission at $\sim$715 nm and a FWHM of 408 nm \citep{Jordi:2006aa}}. The nominal bright limit of $G=6$ was recently overcome and it is now expected that \emph{Gaia} astrometry will be complete at the bright end \citep{MartinFleitas2014}. The spinning satellite's scanning law results in an average number of 70 measurements per source over the five-year nominal mission lifetime, where we identify a measurement as the result of one {pass} in the instrument's focal plane. The exact number of measurements varies as a function of sky position and is highest for objects located at 45\degr\ ecliptic longitude.

The responsibility for processing the \emph{Gaia} telemetry lies with ESA and the Data Processing and Analysis Consortium (DPAC), where the latter consists of $\sim$400 European scientists and is responsible for producing the science data. The processing chain is complex \citep{DPAC:2007kx} and we highlight here {only} the concepts relevant for our study. After a global astrometric solution has been obtained from a reference star sample, the astrometric motions of other stars are modelled with the standard astrometric model consisting of positions, parallax, and proper motions. If excess noise in the residuals are detected and sufficient data {are} present, a sequence of increasingly complex models that include acceleration terms and/or orbital motion are probed until a satisfactory solution is found.

Many nearby binary stars will be solved as astrometric and/or radial velocity binaries by the \emph{Gaia} pipeline. Others will be detected in radial velocity only, because the photocentre motion of the binary is diluted. The purpose of our study is to demonstrate that for some \emph{Gaia} binaries, an additional astrometric signature will be detectable that is caused by a circumbinary planet.

\subsection{Orbit detection}\label{sec:orbdet}
In most cases, the dominant orbital motion will originate from the binary whose period we designate $P_{\rm bin}$. We assume that this motion can me modelled to the level of the \emph{Gaia} single-{measurement} precision $\sigma_{\rm m}$. The signature of the circumbinary planet will then become apparent as an additional periodic term in the residuals of the binary model that has the planet's orbital period $P_{\rm p}$.

\begin{figure}
\center
\includegraphics[width= \linewidth]{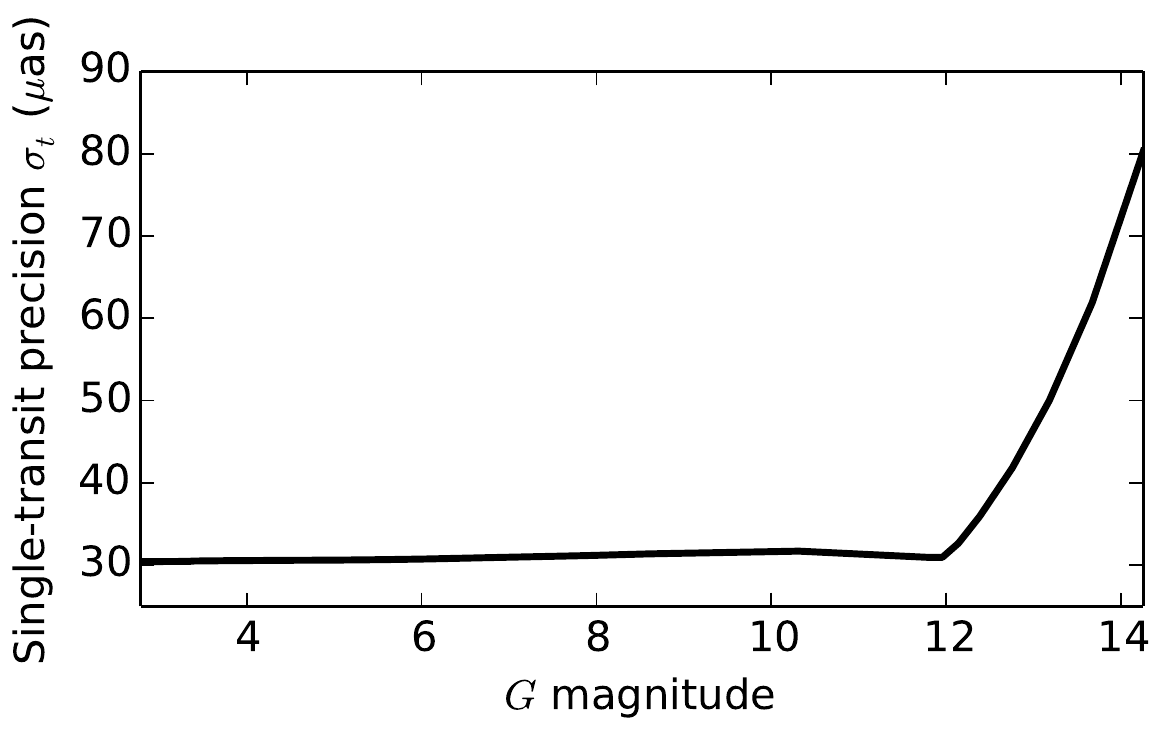}
\caption{Simplified estimate of the \emph{Gaia} single-{measurement} precision as a function of magnitude, following \citet{de-Bruijne:2014aa}.}
\label{fig:sigma_t}
\end{figure}

The two most important parameters for the detection of orbital motion are the number of {measurements} $N_{\rm m}$ and the measurement precision $\sigma_{\rm m}$. We either assumed the average number of 70 {measurement}s over five years or we obtained a refined estimate using DPAC's \emph{Gaia} Observation Schedule Tool\footnote{{http://gaia.esac.esa.int/tomcat\_gost/gost/index.jsp?extra}.}, which yields predicted {observation} times as a function of user-provided object identifiers or target coordinates. \emph{Gaia}'s astrometric precision is a complex function of source magnitude and we used the prescription of \cite{de-Bruijne:2014aa}, which yields the along-scan uncertainty $\sigma_{\rm m}$ as a function of apparent \emph{Gaia} magnitude $G$ and $V-I$ colour\footnote{See also {http://www.cosmos.esa.int/web/gaia/science-performance}.}. For sources brighter than 12th magnitude, the precision is approximately 30 $\mu$as, see Fig. \ref{fig:sigma_t}. This picture neglects the CCD gating scheme that modulates the precision for $G<12$ sources, yet it is sufficient for our purposes.

{The scaling between a binary's photocentre and barycentre orbit size is determined by the magnitude difference between the components and their mass ratio \citep{Heintz:1978yq, Gontcharov:2002aa}. The fractional mass,
\begin{equation}\label{eq:1}
f = M_2 / (M_1+M_2),
\end{equation}
and the fractional luminosity in a certain passband,
\begin{equation}\label{eq:2}
\beta = L_2 / (L_1+L_2) = (1+10^{0.4\ \Delta m})^{-1},
\end{equation}
where $\Delta m$ is the magnitude difference, help to define the relationship between the semimajor axis $\alpha$ of the photocentre orbit and the semimajor axis $a_{\rm rel}$ of the relative orbit, where both are measured in mas:
\begin{equation}\label{eq:3}
\alpha = a_{\rm rel} \, (f-\beta).
\end{equation}
The relative semimajor axis is related to the component masses and the orbital period through Kepler's law:
\begin{equation}\label{eq:4}
G\, (M_1+M_2) = 4\, \pi^2 \frac{\bar a_{\rm rel}^3}{P^2},
\end{equation}
were G is the gravitational constant, $\bar a_{\rm rel}$ is measured in metres and $P$ is in seconds. The relation between $\bar a_{\rm rel}$ and $a_{\rm rel}$ is given by the parallax.} These equations define the astrometric signal for the central binary, where $M_1$ and $M_2$ are the binary masses and $P$ is the binary period $P_\mathrm{bin}$, and of the binary motion caused by the planet, where the central mass is $M_1+M_2$ and the period is the planet period $P_\mathrm{p}$.

We use the astrometric signal-to-noise ratio defined as 
\begin{equation}\label{eq:snr}
S/N = \alpha\, \sqrt{N_{\rm m}} / \sigma_{\rm m}
\end{equation}
to define the threshold for detecting an orbit with a semimajor axis $\alpha$ of the photocentre's orbit. The relationships between orbital parameters, distance, component masses, and signal amplitude can be found, e.g., in \citet{Hilditch:2001uq} and \citet{Sahlmann:2012fk2}. It was shown that $S/N \gtrsim20$ allows for the robust detection of an astrometric orbit, which includes the determination of all orbital parameters with reasonable uncertainties \citep{Catanzarite:2006lr, Neveu:2012fk}. More detail on the choice of this threshold is given in Appendix \ref{sec:app1}. Finally, \emph{Gaia}'s detection capability will be slightly influenced by the actual sampling of the orbital motion and by the orientation distribution scan angles along which the essentially one-dimensional measurement will be made. Due to the way \emph{Gaia} scans the sky, typically {two\,--\,four observations} will be grouped within 6\,--\,12 hours equivalent to one\,--\,two revolutions of the spinning satellite. We neglected the influence of sampling and scan orientation for the sake of simplicity and computation speed.

An additional complication concerning detectability of circumbinary planets stems from the presence of three signatures: (a) the parallax and proper motion of the binary, (b) the orbital motion of the binary with amplitude $\alpha$, and (c) the orbital motion of the binary caused by the planet with amplitude $a_{\rm p,1}$, where $a_{\rm p,1}$ is the barycentric orbit size, because in the vast majority of cases the light contribution by the planet is negligible. These three signals have a total of 19 free parameters and will have to be disentangled. 

As we will show, this should not be problematic in the majority of cases, because there are 70 data points on average and the parallax and the binary orbital motion are detected at much higher $S/N$ than the planet signature.

In addition, circumbinary planets are not expected to exist at periods shorter than the stability limit $P_{\rm crit}\sim 4.5 \, P_{\rm bin}$ (see \citet{Holman:1999lr} for a more thorough criterion), which results in a natural period separation of binary and planet signals. {A simulated example of the orbital photocentre motion of a planet-hosting binary is shown in Fig. \ref{fig:CBP_signature}.}

\subsection{Example 1: Kepler-16}\label{sec:examples}

To showcase the principles of circumbinary planet detection with astrometry, we discuss the Kepler-16 system \citep{Doyle:2011vn}. It consists of a 41.1-day binary with component masses of $M_1=0.6897\,M_{\sun}$ and $M_2=0.2026\,M_{\sun}$, orbited by a planet with mass $M_{\rm p}=0.333\,M_{\rm J}$ in a 228.8-day orbit. The Kepler magnitude of the system is 11.7 and its distance was estimated at $\sim$61 pc.

\subsubsection{Astrometric signature of the binary}
The astrometric semimajor axis of the primary is $a_{\rm bin,1} = 834\, \mu$as and the relative semimajor axis is $a_{\rm bin}=3.7$ mas. Thus the components are not resolved by \emph{Gaia}. Using the binary flux ratio of 0.0155 measured by Kepler, we find that the photocentric semimajor axis is $\alpha = 778\, \mu$as. Over five years, \emph{Gaia} will observe Kepler-16 about 89 times, where we included a $-$10 \% margin to account for dead time that the {observation time} predictor does not incorporate, with an uncertainty of $\sim$30 $\mu$as. The binary's astrometric motion {in} Kepler-16 will therefore be detected {by} \emph{Gaia} astrometry with $S/N\simeq240$. 

\subsubsection{Astrometric signature of the planet}
\begin{figure}
\center
\includegraphics[width= \linewidth]{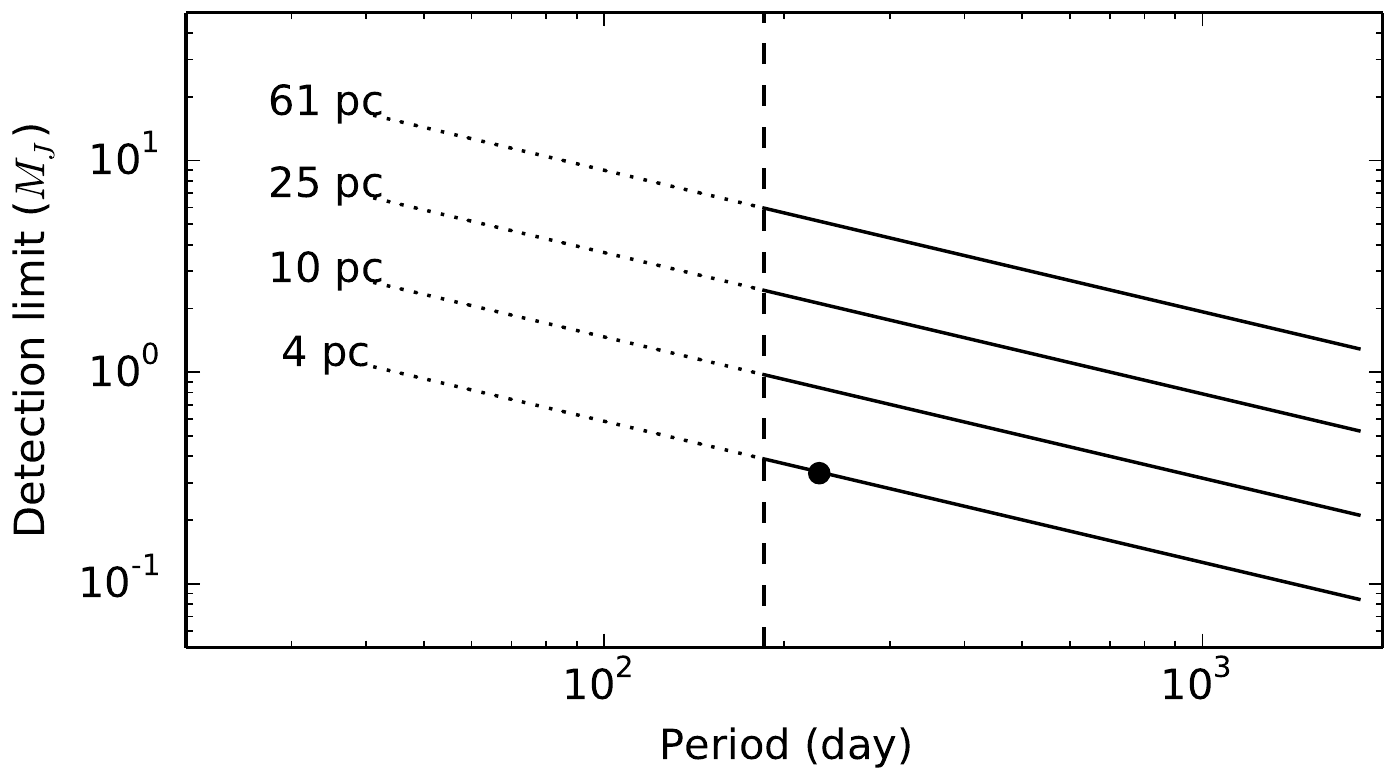}
\caption{Estimated detection limits for planets around Kepler-16 using \emph{Gaia} astrometry. The system is set to distances of 61, 25, 10, and 4 pc and planets above the respective curves would be detected. The vertical dashed line is located at the stability limit $P_{\rm crit}\sim 4.5 \, P_{\rm bin}$, i.e. planets are thought to {be restricted to} the right of this line. The circle marks the location of the planet Kepler-16b, whose orbital signature would be detectable only if Kepler-16 {were} located at a distance of 4 pc.} 
\label{fig:Kepler16_detLimit}
\end{figure}

The planet's gravitational pull will displace the system's barycentre with a semimajor axis of  $a_{\rm p,1} = 4.1\, \mu$as. As shown in Fig. \ref{fig:Kepler16_detLimit}, this is too small to be detected by \emph{Gaia} at a distance of 61 pc. Yet, Fig. \ref{fig:Kepler16_detLimit} illustrates the discovery potential of \emph{Gaia} for circumbinary planets. If a Kepler-16 - like {binary were} located at 10 pc, \emph{Gaia} { would detect} all {orbiting} planets more massive than 1 $M_{\rm J}$ in orbits longer than $P_{\rm crit}\sim188$ days. 

At the actual distance, {\it Gaia} will be able to place a constraint on the existence of a second planet in the Kepler-16 system with a mass larger than two to five times the mass of Jupiter depending on its period.

\subsection{Example 2: Nearby spectroscopic binaries}

We examined the sample of 89 spectroscopic binaries studied by \cite{Halbwachs:2003kx}. An analysis of the {\it Kepler} results shows {a tendency for circumbinary planets to exist in orbits near the inner stability limit, with an over-density at} $\sim$$ 6 \,P_{\rm bin}$ \citep{Orosz:2012aa, Martin:2014aa}. We selected a subsample only including the 38 binaries that have periods in the range 10 days $< P_{\rm bin} < $ {304 days} ($=5/6$ years), to ensure that a putative planet would complete a revolution around its binary host {within the nominal five-year \emph{Gaia} mission. }

Distances to these systems were retrieved from the Hipparcos catalogue \citep{ESA:1997vn} or from \cite{van-Altena:1995aa} (GJ 92.1, GJ 423 A, GJ 433.2B, GJ 615.2A) and \cite{Crissman:1957aa} (GJ 1064 B). The distances to objects in the Pleiades and Praesepe were assumed constant at 120 pc (e.g.\ \citealt{Palmer:2014aa}) and 182 pc (e.g.\ \citealt{Boudreault:2012aa}), respectively. Orbital parameters, primary masses, and mass ratios were taken from \cite{Halbwachs:2003kx}. When only the minimum mass ratio $q_{\rm min}$ was known, we {used} an estimate of $q = 1.15 \,q_{\rm min}$, which corresponds to the median correction in a sample of randomly oriented orbits (see, e.g., \citealt{Sahlmann:2012fk2}). When only the maximum mass ratio $q_{\rm max}$ was known, we assumed $q = q_{\rm max}$.

\begin{figure}
\center
\includegraphics[width= \linewidth]{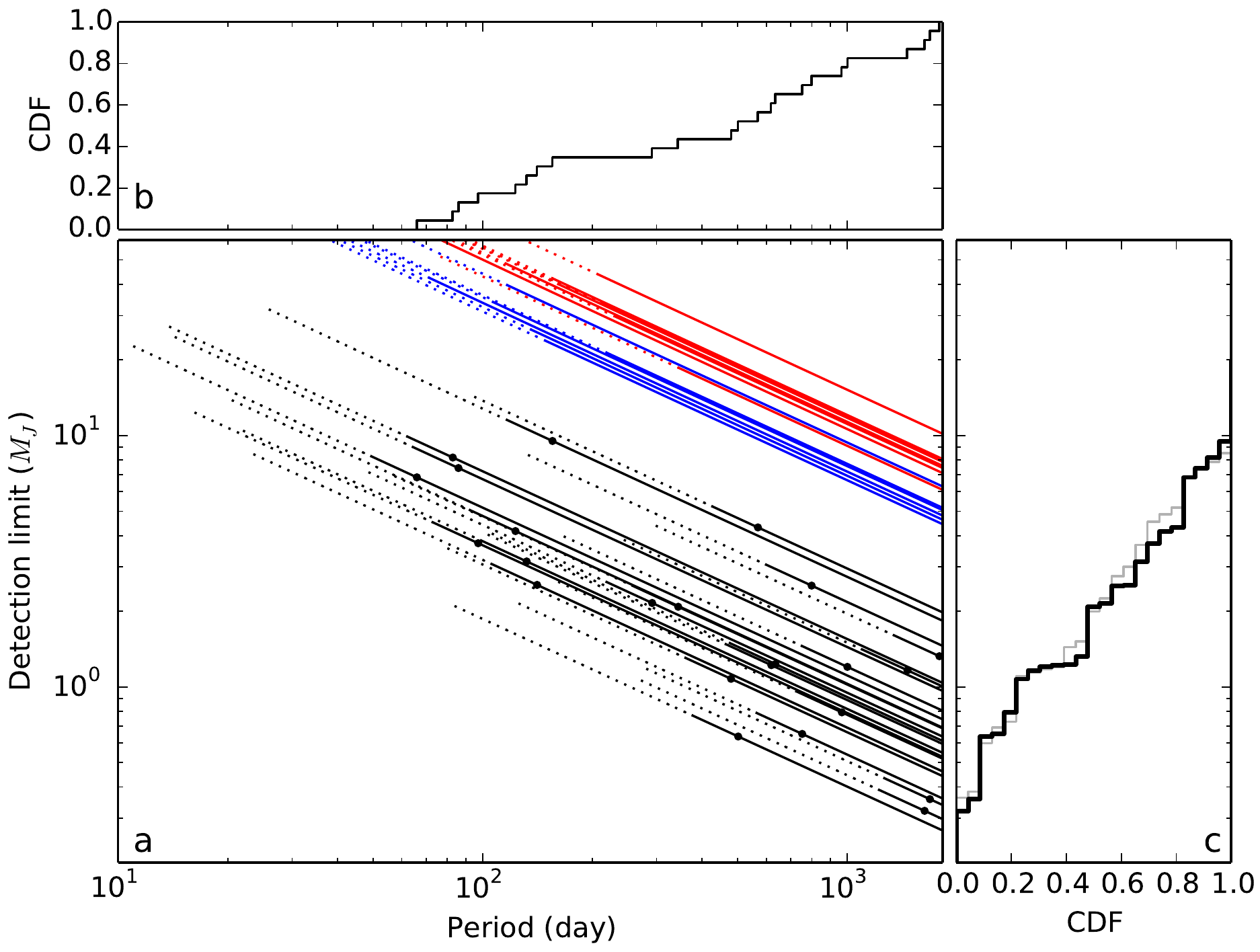}
\caption{Panel \textbf{a}: Estimated detection limits for planets around 38 spectroscopic binaries having 10 days $< P_{\rm bin} < 304$ days. The display is similar to Fig. \ref{fig:Kepler16_detLimit}. Red and blue curves correspond to binaries in Praesepe (9 objects) and the Pleiades (6 objects), respectively. The black curves correspond to the 23 field binaries, which are at distances $<$ 52 pc. {All curves terminate at the binary period, and appear as dotted lines for $P_{\rm p} < 4.5P_{\rm bin}$}. Dots mark the location of hypothetical planets with $P_{\rm p} = 6\,P_{\rm bin}$. Panels \textbf{b} and \textbf{c} show cumulative histograms of these planets.}
\label{fig:Halbwachs_detLimit}
\end{figure}

Figure \ref{fig:Halbwachs_detLimit} shows the planet detection limits for these systems. For about half of the field binaries, \emph{Gaia} could detect Jupiter-mass planets (and lighter) with orbital periods close to the nominal mission lifetime. For five binaries, \emph{Gaia} could detect sub-Jupiter-mass planets with $P_{\rm p} = 6\,P_{\rm bin}$, which corresponds to the {currently observed} pile-up of Kepler circumbinary planets.

We performed these calculations with the average number of $N_t=70$ {measurements} (grey curve in Fig. \ref{fig:Halbwachs_detLimit}\textbf{c}) and with a more accurate estimate obtained from the {observation time} predictor (black curve in Fig. \ref{fig:Halbwachs_detLimit}\textbf{c}), again accounting for a 10 \% margin. Both cases lead essentially to the same conclusion.

The 23 field binaries are located closer than 52 pc, {meaning that} their parallaxes will be detected with $S/N\gtrsim5500$. 18 of those have relative separations smaller than 30 mas, i.e.\ will hardly be resolved by \emph{Gaia}\footnote{The short side of one of \emph{Gaia}'s rectangular pixels measures $\sim$59 mas on the sky and corresponds to approximately 1 FWHM of the short-wavelength PSF.} and their binary photocentric motions will be detected with $S/N=70-1900$. Importantly, if a planet is detected orbiting any of these binaries, the {3-d} mutual inclination between the binary and planetary orbital planes can be calculated with an accuracy of $\sim$$10\degr$, similar to what is achieved for the spin-orbit obliquity\footnote{{Here, obliquity denotes the projected spin-orbit angle.}} by observing the Rossiter-McLaughlin effect of single stars (e.g. \citealt{Lendl:2014aa}).

For the binaries in clusters, which are at larger distances, \emph{Gaia} will be able to detect the most massive circumbinary planets in the range of $5-30\,M_{\rm J}$ and with orbital periods shorter than the mission lifetime. 

This shows that \emph{Gaia} will set constraints on the presence of planets around many nearby known binaries. Since \emph{Gaia} is an all-sky survey, many nearby binaries are expected to be newly discovered. This will result in a large sample of binaries and renders the detection of circumbinary planets with \emph{Gaia} astrometry very likely.

\section{Estimation of \emph{Gaia}'s circumbinary planet discovery yield within 200 pc}\label{sec:simulation}
To estimate the number of circumbinary planets \emph{Gaia} may detect, we simulated the population of binary stars with FGK-dwarf primaries in the solar neighbourhood. We set the horizon at 200 pc distance to be able to extrapolate the `locally` measured properties with high confidence. The principal uncertainty originates in the {true} occurrence {and distribution} of circumbinary planets, which are unknown. We extrapolated the planet properties from their better known population around single FGK-dwarfs. This approach is analogous to the work of \cite{Martin:2014aa}, yet adapted to the specificities of astrometric planet detection. 

\subsection{Binary sample}
Our synthetic binary sample is constructed from a combination of the results from {\it Kepler} and from radial velocity surveys, using a methodology similar to \citet{Martin:2014aa} that we remind briefly here. We first select stars in the {\it Kepler} target catalog with {\it Data Availability Flag = 2}, meaning that the star has been logged and observed, and with measured effective temperatures, surface gravities and metallicities. We then use a relation from \citet{Torres:2010uq} to estimate masses for each star. The list is cut to only include stars with masses between 0.6 and 1.3$M_{\sun}$, so that it coincides with the Halbwachs survey. 

These stars are made the primary stars in binaries constructed using the results from radial velocity surveys. The secondary masses come from the mass ratio distribution in \citet{Halbwachs:2003kx}. The binary periods are drawn from the log-normal distribution calculated in \citet{Duquennoy:1991kx}, and restricted to be within 1 day and 10 years, to correspond to the Halbwachs survey. The inclination of the binary on the sky was randomised according to a uniform distribution in $\cos I_{\rm bin}$.

We compute absolute $V$ magnitude from stellar masses with the relationships of \cite{Henry:1993aa} and obtain $V-I$ colours on the main-sequence from \cite{Cox:2000uq}. The absolute $G$ magnitude is obtained as in \cite{Jordi:2010kx} and \cite{de-Bruijne:2014aa}. Given a distance, we can obtain all relevant parameters of a binary system, in particular the systems apparent magnitude, the relative orbit size, and the semimajor axis $\alpha$ of the photocentric binary orbit.

\subsection{Simulated binaries in the solar neighbourhood}

We assumed a constant space density of FGK dwarfs within a 200 pc sphere around the Sun on the basis of the RECONS  number count \citep{ESA:1997vn, Henry:2006uq}. There are 70 FGK dwarfs within 10 pc of the Sun\footnote{{http://www.recons.org/census.posted.htm} in August 2014.}, which corresponds to a density of 0.0167 pc$^{-3}$. The fraction of main-sequence stars in multiple systems with periods $<$10 years was determined to $13.5^{+1.8}_{-1.6}$ \% \citep{Halbwachs:2003kx}, which is the rate we used. 

Because the astrometric planet detection capability depends critically on the target's distance from the Sun, we generated star populations in spherical shells with equal distance spacing, i.e. spanning ranges of 0--5 pc, 5--10 pc, 10--15 pc and so forth. Every shell was populated with $N_{\star}$ stars on the basis of its volume (Table \ref{tab:1}) and the number of binaries $N_{\rm bin}$ was calculated. The properties of these binaries were determined by parameter distributions of randomly selected binary systems from the {synthetic} binary sample. 

\subsection{Simulated {giant} circumbinary planets}
\begin{figure}
\center
\includegraphics[width= \linewidth]{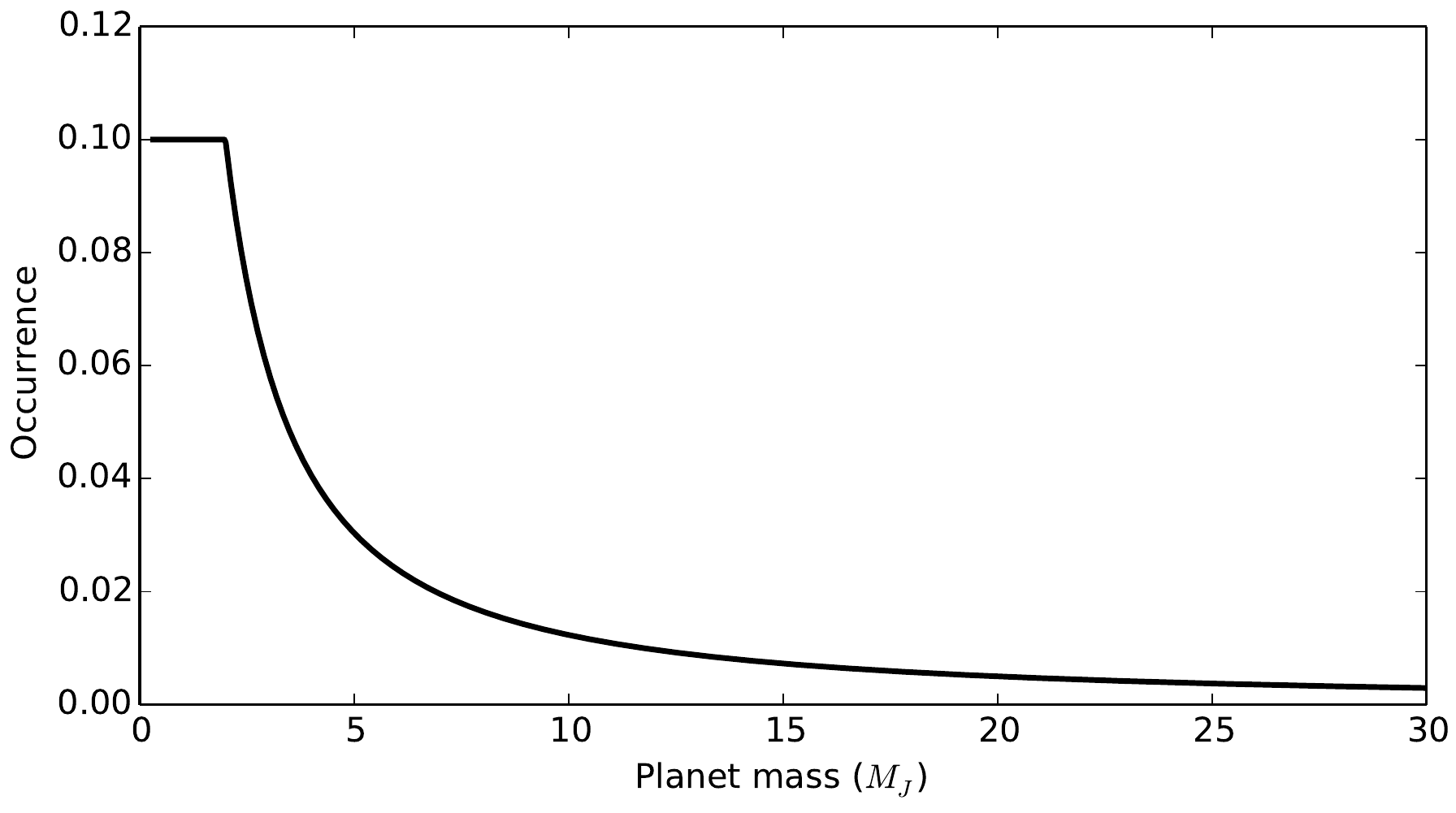}
\caption{The mass distribution of giant planets used in the simulation.}
\label{fig:inputPlanetMassDist}
\end{figure}

We assumed a constant rate of giant planet occurrence around binary stars of 10 \% and we supposed that every planet is located at the stability limit pile-up of $6\,P_{\rm bin}$ {with one planet per binary star}\footnote{{Placing all planets at the pile-up period results in a lower limit on the detectability, because further out induce a larger astrometric signal, thus are easier to detected as long as their period is covered by the measurement timespan.}}. To draw planet masses, we synthesised a mass-dependent relative abundance of giant planets in the range of 0.31 -- 30 $M_{\rm J}$  from the results of radial velocity surveys \citep{Segransan:2010xr, Mayor:2011fj, Sahlmann:2011fk}, see Fig. \ref{fig:inputPlanetMassDist}. {Our simulations are thus strictly limited to giant planets more massive than Saturn.} For every simulated binary hosting a planet, the planet mass is determined via the probability density function (PDF) $f_{\rm p, mass}$ defined by the normalised version of this distribution. We discuss the effects of altering these assumptions in Sect \ref{sec:revisit}.

\subsection{The population of {detectable} planets} 
We set up our simulation to process a sequence of spherical shells. For every shell we obtain a set of synthesised binary systems and we discard binaries with periods longer than 304 days, because we impose that planets can only be detected if one complete orbital period is observed. The remaining binaries are conservatively set to the distance of the outer shell limit and are assigned planets according to the 10 \% occurrence rate. 

For every binary with a planet we compute the astrometric detection mass-limit $M_{\rm p,lim}$ for a planet orbiting at the pile-up period $P_{\rm p} = 6 \,P_{\rm bin}$. The mass detection-limit takes into account the total mass of the binary, its combined $G$ magnitude and the corresponding single-{measurement} precision, and the requirement of $S/N>20$, where we used $N_{\rm m}=70$. The probability $r_{\rm det}$ of detecting this planet corresponds to the integral of $f_{\rm p,mass}$ over masses larger than the detection limit $M_{\rm p,lim}$. By averaging this probability for all binaries, we obtain the mean probability $\bar r_{\rm det}$ of a planet detection in every shell. The number of detected circumbinary planets is obtained from $N_{\rm CBP}= N_{\rm p} \cdot \bar r_{\rm det}$, where the meaning of $N_{\rm p}$ and other variables is summarised in Table~\ref{tab:2}.

\begin{table}
\caption{Relevant parameters}
\label{tab:2}
\begin{tabular}{cllrrrrr}
\hline
Symbol&Description&Value\\
\hline
$N_{\star}$ & Number of stars in shell $s$ & Table \ref{tab:1}\\[1pt]
$r_{\rm bin}$ & Binary fraction & 13.9 \% \\[1pt]
$r_{\rm pl}$ & Fraction of binaries with one planet & 10 \% \\[1pt]
$N_{\rm bin,P}$ & \parbox[t]{5cm}{Number of accessible binaries ($P_\mathrm{bin}<304$ days)}& Table \ref{tab:1}\\ 
$N_{\rm p}$ & \parbox[t]{5cm}{Number of planets around\\accessible binaries} & Table \ref{tab:1}\\
$\bar r_{\rm det}$ & \parbox[t]{5cm}{Average probability that\\a giant planet is detectable} & Table \ref{tab:1}\\
$N_{\rm CBP}$ & Number of planets detected in shell $s$ & Table \ref{tab:1}\\
$r_{\rm res}$ & Rate of potentially resolved binaries & Table \ref{tab:1}\\
$r_{S/N}$ & Rate of binaries with $S/N>100$ & Table \ref{tab:1}\\
\hline
\end{tabular}
\end{table}

\begin{table}
\setlength{\tabcolsep}{4pt}
\caption{Results of the simulation}
\label{tab:1}
\small
\begin{tabular}{rrrrrrrr}
\hline
Shell&$N_{\star}$ & $N_{\rm bin,P}$ & $N_{\rm p}$ & $\bar r_{\rm det}$ & $N_{\rm CBP}$ & $r_{\rm res}$& $r_{ S/N}$\\
(pc)& & && & &&   \\
\hline
$  0-  5$ &   9 &   1 & 0.1 & 0.857 & 0.0 & 0.21 & 1.0 \\
$  5- 10$ &  61 &   4 & 0.4 & 0.743 & 0.3 & 0.13 & 0.9 \\
$ 10- 15$ & 166 &  11 & 1.1 & 0.651 & 0.7 & 0.09 & 0.9 \\
$ 15- 20$ & 324 &  22 & 2.2 & 0.587 & 1.3 & 0.06 & 0.9 \\
$ 20- 25$ & 534 &  36 & 3.6 & 0.527 & 1.9 & 0.04 & 0.8 \\
$ 25- 30$ & 796 &  53 & 5.3 & 0.475 & 2.5 & 0.02 & 0.8 \\
$ 30- 35$ & 1111 &  75 & 7.5 & 0.438 & 3.3 & 0.01 & 0.8 \\
$ 35- 40$ & 1479 & 101 & 10.1 & 0.402 & 4.0 & 0.00 & 0.8 \\
$ 40- 45$ & 1899 & 130 & 13.0 & 0.379 & 4.9 & 0.00 & 0.7 \\
$ 45- 50$ & 2371 & 160 & 16.0 & 0.353 & 5.7 & 0.00 & 0.7 \\
$ 50- 55$ & 2896 & 195 & 19.5 & 0.329 & 6.4 & 0.00 & 0.7 \\
$ 55- 60$ & 3474 & 236 & 23.6 & 0.301 & 7.1 & 0.00 & 0.7 \\
$ 60- 65$ & 4104 & 277 & 27.7 & 0.285 & 7.9 & 0.00 & 0.7 \\
$ 65- 70$ & 4786 & 324 & 32.4 & 0.275 & 8.9 & 0.00 & 0.7 \\
$ 70- 75$ & 5521 & 375 & 37.5 & 0.257 & 9.7 & 0.00 & 0.6 \\
$ 75- 80$ & 6309 & 421 & 42.1 & 0.245 & 10.3 & 0.00 & 0.6 \\
$ 80- 85$ & 7149 & 483 & 48.3 & 0.232 & 11.2 & 0.00 & 0.6 \\
$ 85- 90$ & 8041 & 550 & 55.0 & 0.222 & 12.2 & 0.00 & 0.6 \\
$ 90- 95$ & 8986 & 602 & 60.2 & 0.207 & 12.4 & 0.00 & 0.6 \\
$ 95-100$ & 9984 & 675 & 67.5 & 0.200 & 13.5 & 0.00 & 0.6 \\
$100-105$ & 11034 & 751 & 75.1 & 0.187 & 14.0 & 0.00 & 0.6 \\
$105-110$ & 12136 & 812 & 81.2 & 0.186 & 15.1 & 0.00 & 0.6 \\
$110-115$ & 13291 & 888 & 88.8 & 0.177 & 15.8 & 0.00 & 0.5 \\
$115-120$ & 14499 & 953 & 95.3 & 0.167 & 15.9 & 0.00 & 0.5 \\
$120-125$ & 15759 & 1071 & 107.1 & 0.160 & 17.2 & 0.00 & 0.5 \\
$125-130$ & 17071 & 1148 & 114.8 & 0.155 & 17.8 & 0.00 & 0.5 \\
$130-135$ & 18436 & 1264 & 126.4 & 0.147 & 18.5 & 0.00 & 0.5 \\
$135-140$ & 19854 & 1330 & 133.0 & 0.138 & 18.4 & 0.00 & 0.5 \\
$140-145$ & 21324 & 1444 & 144.4 & 0.134 & 19.3 & 0.00 & 0.5 \\
$145-150$ & 22846 & 1538 & 153.8 & 0.133 & 20.5 & 0.00 & 0.5 \\
$150-155$ & 24421 & 1665 & 166.5 & 0.121 & 20.1 & 0.00 & 0.4 \\
$155-160$ & 26049 & 1767 & 176.7 & 0.121 & 21.3 & 0.00 & 0.5 \\
$160-165$ & 27729 & 1874 & 187.4 & 0.111 & 20.8 & 0.00 & 0.4 \\
$165-170$ & 29461 & 1967 & 196.7 & 0.109 & 21.4 & 0.00 & 0.4 \\
$170-175$ & 31246 & 2121 & 212.1 & 0.105 & 22.2 & 0.00 & 0.4 \\
$175-180$ & 33084 & 2202 & 220.2 & 0.100 & 22.0 & 0.00 & 0.4 \\
$180-185$ & 34974 & 2314 & 231.4 & 0.096 & 22.1 & 0.00 & 0.4 \\
$185-190$ & 36916 & 2501 & 250.1 & 0.090 & 22.6 & 0.00 & 0.4 \\
$190-195$ & 38911 & 2627 & 262.7 & 0.090 & 23.5 & 0.00 & 0.4 \\
$195-200$ & 40959 & 2725 & 272.5 & 0.084 & 22.9 & 0.00 & 0.4 \\
Total & $5.6\!\cdot\!10^5$ & 37691 & 3769 & -- & 516 & -- & -- \\

\hline
\end{tabular}
\end{table}

\begin{figure}
\center
\includegraphics[width= \linewidth]{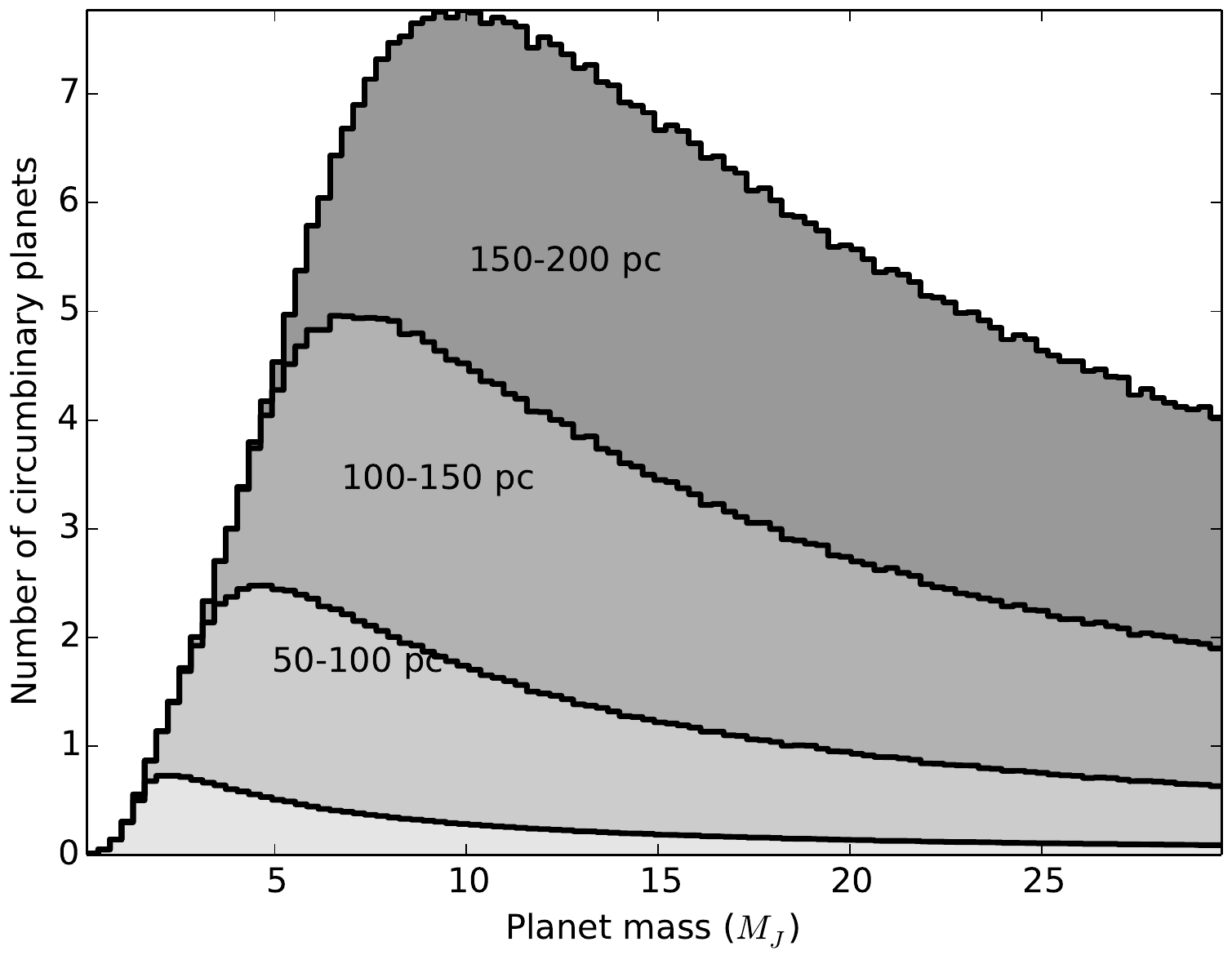}
\caption{Mass histogram of {detectable} circumbinary planets. The four curves correspond to planets detected within spheres of $\leqslant50$ pc (25 planets), $\leqslant100$ pc (124 planets), $\leqslant150$ pc (297 planets), and $\leqslant200$ pc (516 planets).}
\label{fig:histomass}
\end{figure}
\begin{figure}
\center
\includegraphics[width= \linewidth]{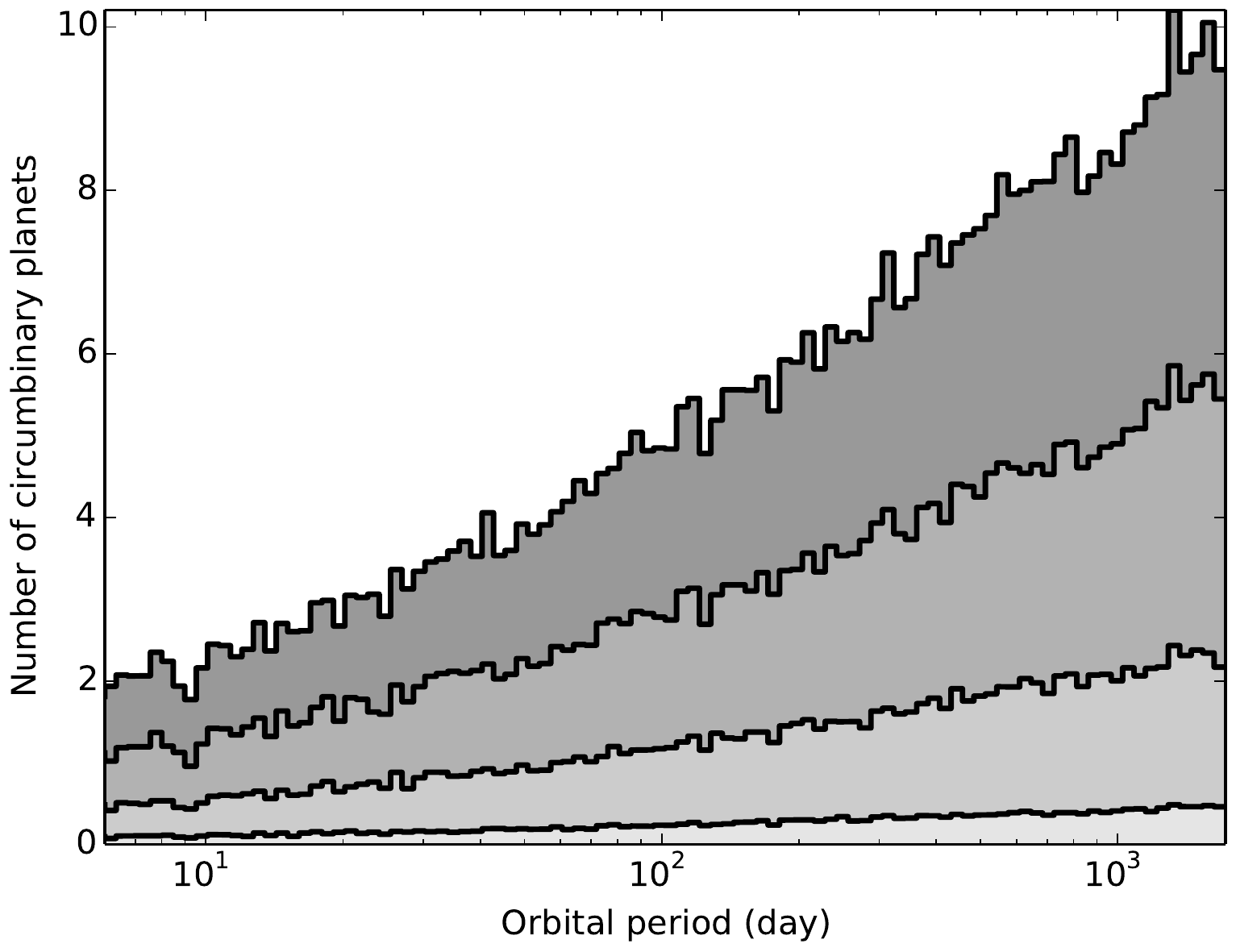}
\caption{Period histogram of {detectable} circumbinary planets. By construction, the distribution of binary periods follows the same {distribution} function, {shifted to} one sixth of the period. The greyscale coding is like in Fig. \ref{fig:histomass}.}
\label{fig:histoperiod}
\end{figure}

The simulation results are summarised in Table~\ref{tab:1}. Figures \ref{fig:histomass} and \ref{fig:histoperiod} show the histograms of masses and periods of detected planets and Figs. \ref{fig:histomasscum} and \ref{fig:histoperiodcum} show their cumulative versions.

In this simulation we find that \emph{Gaia} will discover 516 planets orbiting binary stars within 200 pc. Within spheres of $\leqslant50$ pc, $\leqslant100$ pc, and $\leqslant150$ pc around the Sun, we predict 25, 124, and 297 planet detections, respectively. Before discussing these absolute numbers, we can observe the characteristics of the planet population. The probability $\bar r_{\rm det}$ of detecting a planet decreases with the distance of the shell. This is because the sky-projected astrometric orbit size {becomes smaller, hence the minimum detectable planet mass increases}. Yet, the number of detected planets increases with distance, because the number of binaries increases as distance to the third power. Most planets are found at long periods, because the astrometric signal increases with orbital period. Finally, we see that the large majority of detected planets have masses $>5\,M_{\rm J}$. Although those are less frequent around binaries (Fig. \ref{fig:inputPlanetMassDist}), they can be detected out to large distances and thus around many more binary stars. 

Table \ref{tab:1} also lists two parameters related to the binaries themselves. Complications in the photocentre measurement process can arise when binaries are visually resolved in the \emph{Gaia} focal plane. We assume that this can be the case if the projected separation of the binary components is larger than 30 mas and their magnitude difference $\Delta G$ is smaller than two magnitudes. The rate of such binaries is indicated by the $r_{\rm res}$ parameter and is smaller than 10 \% and 1 \% for binaries beyond 20 pc and 40 pc, respectively. Therefore, resolved binaries will have to be considered only for very few, very nearby circumbinary discoveries.

The other binary parameter $r_{S/N}$ represents the rate of binaries whose photocentric motions are detected with $S/N>100$. This rate reaches unity for the most nearby systems and decreases slowly to 0.5 at 200 pc. This means that most of the binary motions will be detected with $S/N>100$. In addition, the measurement timespan covers at least 6 orbital periods, thus the accurate determination of the binary parameters is almost guaranteed. The parallax of most FGK-dwarf binaries within 200 pc will be measured with $S/N \gtrsim 1400$. This validates what we claimed in Sect. \ref{sec:orbdet}: Since the two dominant signals, the parallax and the binary orbit, will be detected with very high $S/N$, it will be generally possible to disentangle the planetary orbit at $S/N>20$, even with uneven time sampling and 19 free parameters constrained by 70 independent measurements.

\subsection{Revisiting our assumptions}\label{sec:revisit}
We quantify the effect of altering our initial assumptions by repeating the simulation and changing the main parameters one by one.

\begin{itemize}
  \item The density of FGK dwarfs, the binary rate, and the circumbinary planet rate are proportional factors in the sense that changing them will influence the number of detected planets directly. If the planet rate is only 5 \%, the number of detected planets is reduced to 258, i.e. half of the initial count.
  \item Period of the planet: It is unlikely that all planets orbit with periods of $6\, P_{\rm bin}$ and we therefore repeated simulations with values of $8\, P_{\rm bin}$ and $10\, P_{\rm bin}$, yielding 479 (93 \%) and 454 (88 \%) detections, respectively. The decrease is attributable to the smaller number of accessible binaries when restricting their period range due to the requirement of covering one planet period with \emph{Gaia} data.
  \item Single-{measurement} precision and number of {measurement}s: Both parameters influence the detection criterion in $S/N$. We found that degrading the precision by 20 \% yields 422 (82 \%) planets and assuming a smaller number of 60 {measurement}s over 5 years results in 474 (92 \%) planet detections.
  \item $S/N$ threshold: As shown in Fig. \ref{fig:HIPSNR}, some orbital signals can be detected with $S/N=15$. Lowering our detection threshold to this value results in 689 planets (134 \%), whereas increasing it to $S/N>30$ yields 317 planets (62 \%). This parameter will determine the false positive rate of the planet discovery.
 \item The planet mass distribution is the most sensitive parameter in our simulation. We have initially assumed that the mass distribution is the same for giant planets around single stars and around binaries. If circumbinary giant planets do not exceed masses of 2 $M_{\rm J}$\footnote{ At a recent conference in Litoms\v yl, Czech Republic, Welsh et al. presented three circumbinary candidates discovered using the ETV method, one of which has a mass of 2.6 $M_{\rm J}$.}, we predict that \emph{Gaia} will discover only four of them (0.807 \%). Such planets can be discovered only around the most nearby binaries, which are few in number.
 
This sensitivity to planet mass means that \emph{Gaia} will disclose the mass distribution of giant planets around binaries, thus making new insights into planet formation processes possible. 
\item \emph{Gaia} mission duration: Our simulations assumed the nominal mission duration of five years. In the hypothetical case that the mission could be extended to a lifetime of 10 years, the number of circumbinary planets increases by a factor of $\sim$2.5 to 1249. The higher than proportional gain is caused by three factors: (1) the higher number of measurements leads to improved detection limits; (2) a longer mission allows to detect planets of longer orbital period, i.e. the number of accessible binary stars is also larger; (3) a longer mission allows to detect planets of lower mass, because the astrometric signal increases with orbital period.
\item Incomplete orbit coverage: Our detection criterion required full coverage of the planet period during the mission. The number of detections is thus conservative, because \emph{Gaia} will also detect some planets with incomplete orbit coverage ($\ga50$ \%) from curvature in the binary fit residuals.

\item {Multi-planet systems: We assumed one single planet per binary, which neglected multi-planet systems like Kepler-47 \citep{Orosz:2012ab}. The presence of additional bodies in a system may complicate the data analysis if their signatures are comparable to the measurement precision. Like around single stars, we expect that circumbinary giant planets are rarely found in tightly packed systems. Because the period of a second planet will then be much longer than the simulated inner planet period $6\, P_{\rm bin}$, we expect that such systems may reveal themselves through an additional long-term drift.}

\end{itemize}
In summary, we find that the \emph{Gaia} circumbinary planet yield in terms of the order of magnitude of discoveries is relatively insensitive to the assumed parameters, except for the high-mass tail of the planet mass distribution. Assuming that very massive giant planets ($2-30\,M_{\rm J}$) exist around binaries as they do around single stars, the number of \emph{Gaia} discoveries will be in the range of 200--600. This is one order of magnitude higher than the number of currently known circumbinary planets.

\section{Discussion}\label{sec:discussion}
It is clear that the sheer number of predicted planets around FGK-dwarf binaries discovered by the \emph{Gaia} survey is the main result of this study. Yet, there are some considerations that further emphasize the importance of this work.

\subsection{Mutual inclinations}
Since most of the binaries studied here will have their orbital inclination measured to better than $\sim$1--5\degr, the uncertainty on the mutual inclination will be dominated by the planetary orbit determination (cf. the discussion in Appendix \ref{sec:app1} and Fig. \ref{fig:HIPinc}). On average, we expect the latter to be uncertain at the 10\degr\ level.

We will then have a sample of potentially hundreds of circumbinary planets with measured mutual orbit inclinations, leading to insights into their formation history and dynamical evolution.

\subsection{Combination with auxiliary data}
At distances larger than $\sim$100 pc, the determination of binary orbital parameters can be enhanced by complementing the \emph{Gaia} astrometry with radial velocity data, either from the on-board radial velocity spectrometer or from independent observations. Radial velocity measurements with km/s precision that are quasi-independent of distance within 200 pc help detecting the astrometric binary orbit and leave more degrees of freedom for the detection of the circumbinary planet orbit.

{ For specific systems, }{ground-based high-resolution spectrographs, built for the detection of exoplanets, will complement unclosed orbits and may reveal the precession of planetary orbits inclined with respect to the binary plane through follow-up observations\footnote{{A precessing planet is not following a Keplerian orbit, and hence fitting Keplerians to a radial velocity curve will fail if data with sufficient precision and timespan are available.}}. Those will refine measurements on the mutual inclination.}

\subsection{Direct detection of circumbinary planets}
Our detection criterion imposed that the planet's orbital period is fully covered by the \emph{Gaia} measurement timespan. Consequently the projected relative separation between a binary and its planet is very small, typically $<$25 mas. In our simulation, there is only one planet with a relative separation $>$100 mas (assuming a 10-year mission, we predict {six} planets with separation $>$100 mas). Therefore, most circumbinary planets whose orbits are characterised by \emph{Gaia} will lie beyond the capabilities of present and upcoming instruments. 

However, \emph{Gaia} will also detect nearby binary stars that exhibit nonlinear deviations from proper motion, which can be indicative of an orbiting planet. These are excellent targets for follow-up studies in particular those aiming at direct detection. Because the planets discovered by non-linear motion are at larger separation from their host star, they can be directly detected with coronagraphic or interferometric instruments on the ground, e.g.\ GPI \citep{Macintosh:2014ab} and SPHERE \citep{Beuzit:2006aa}, or in space, e.g.\ JWST/NIRISS \citep{artigau2014}.

\subsection{Planets around non-FGK binaries}
We considered FGK binaries in our study because the planet population around FGK dwarfs is the best studied so far. Yet, we expect that \emph{Gaia} will discover planets around other types of binaries on and off the main sequence. Although the binary fraction drops {as primary mass decreases} \citep{Raghavan:2010lr}, since M dwarfs are 3.5 times more abundant than FGK dwarfs in the solar neighbourhood, there will be plenty of M dwarf binaries amendable for planet search with \emph{Gaia}. On the hotter side, A star binaries could be searched for planets too, with the particular interest of detecting them at younger ages, which is important for an eventual detection using direct imaging.  

{A large sample of circumbinary planets will open a meaningful study of the effect of stellar mass on the presence of planets. {\it Gaia} is not as sensitive to {adverse effects associated to a star's spectral type, e.g.\ photometric and spectroscopic variability or the availability of spectral lines,} as the other detection methods. This means it will explore a wider mass range than the radial-velocity and transit techniques. In addition, the binary nature of the host can become a big help as well: Protoplanetary disc masses scale with the central mass, be it a single or binary star \citep{Andrews:2013uq}, and heavier discs are expected to produce more gas giants \citep{Mordasini:2012ve}. A G dwarf + G dwarf binary, for instance, had a disc mass equivalent to a single A star. }

{Furthermore, {\it Gaia} will contribute to the study of planets affected by the natural evolution of binaries across the H-R diagram, from pre-main sequence to stellar remnants.} There are several claims of circumbinary planet detection around eclipsing binaries and post-common envelope binaries using measurements of eclipse-timing variations. We list the systems with available distance estimation in Table \ref{tab:nnser} and computed approximate astrometric parameters for the proposed planetary solutions and the \emph{Gaia} observations. All planet periods are longer than the five-year nominal \emph{Gaia} mission with fractional orbit coverages of $r_{\rm orb}=0.09-0.66$. The probability of detecting the planet signal depends on $S/N$ and $r_{\rm orb}$. The best candidate for validation with \emph{Gaia} is QS Vir \citep{Qian:2010ab} that has a large predicted signal amplitude of 443 $\mu$as ($S/N=44$) and an orbital period that is almost covered by \emph{Gaia} measurements ($r_{\rm orb}=0.63$).

{Finally, there is no expected limitation of {\it Gaia}'s planet detection capability caused by stellar activity \citep{Eriksson:2007uq}. For a similar total mass, {\it Gaia}'s main constraint is the distance and apparent magnitude of a system. Thanks to this, it will be possible to detect planets from the pre-main sequence to stellar remnants and through the red-giant phase, thereby quantifying the influence of stellar evolution on the abundance, architecture, and mass distribution of planetary systems. This way, {\it Gaia} will confront theories about the consequences of mass loss and about the existence and efficiency of an eventual second epoch of planet formation\footnote{These arguments are applicable to single stars as well, although they have not been explored in the literature.}.}

\subsection{Circumbinary brown dwarfs}
\emph{Gaia}'s sensitivity to circumbinary companions increases rapidly with the companion mass. Brown dwarfs in orbits with periods not much longer than the mission lifetime around nearby binaries will therefore be efficiently discovered and characterised by \emph{Gaia}. Again, this will provide a critical test for the comparison with brown dwarf companions within 10 AU of single stars \citep{Sahlmann:2011qy}, in particular the distribution of masses and orbital parameters.

\section{Conclusion}

Starting from empirical assumptions about the population of binary stars and of gas giants, we built a model of the circumbinary planet population in a volume of 200 pc. The precision of individual astrometric  measurements by {\it Gaia} and the length of the survey inform that of the order of 500 circumbinary gas giants could be detected. The vast majority, however, will have masses in excess of fives times that of Jupiter. If the lack of planets more massive than Jupiter in the {\it Kepler} results reflects an intrinsic property of the circumbinary planet population, {\it Gaia} will produce a stringent null result that can enter planet formation models. In reverse, if those planets do exist {as very recent results suggest}, by the sheer number that will be produced, we will obtain the opportunity to study circumbinary planet properties and abundances in detail, not least to study how those are influenced by the evolution of their hosts. 

The aptitude to deliver mutual inclinations between the binary and planetary orbital planes, will further our understanding of dynamical processes happening post planet-formation, completing the information collected on hot Jupiters thanks to the Rossiter-McLaughlin effect, to long period planets and a wide range of stellar properties.

{According to the mission schedule, the first circumbinary planet discoveries of \emph{Gaia} can be expected in the fourth data release, i.e.\ not before 2018.}

{As a final remark, we note that the 10 \% rate of circumbinary planets that we used is a conservative number. About 20--30 \% of the \emph{Kepler} binaries are in fact fairly tight triple systems \citep{Rappaport:2013aa}, which is compatible with what is found in the 100 pc solar neighbourhood \citep{Tokovinin:2006la}, and \emph{Gaia} will easily identify these triple-star systems. It is likely that tertiary stars, if close enough, will disrupt any circumbinary protoplanetary discs, and hence inhibit planet formation \citep{Mazeh:1979eu}. This effectively increases the planet rate to $\sim$13 \%. Furthermore, new circumbinary planets keep being found with \emph{Kepler}, of which some have misaligned orbits. All these factors augment the expected \emph{Gaia} discovery rate.}

\section*{Acknowledgments}
J.S. is supported by an ESA research fellowship. A. H. M. J. Triaud is a Swiss National Science Foundation fellow under grant number P300P2-147773. D. V. Martin is funded by the Swiss National Science Foundation. This research made use of the databases at the Centre de Donn\'ees astronomiques de Strasbourg ({http: //cds.u-strasbg.fr/}); of NASA's Astrophysics Data System Service ({http://adsabs.harvard.edu/abstract\_service.html}; of the paper repositories at arXiv; and of Astropy, a community-developed core Python package for Astronomy \citep{Astropy-Collaboration:2013aa}.  JS and AT attended the Cambridge symposium on {\it Characterizing planetary systems across the HR diagram}, which inspired some elements of our discussion. We acknowledge the role of office 443 B at Observatoire de Gen\`eve for stimulating this research. {We thank the anonymous referee whose comments helped to improve the presentation of our results. JS thanks the members of the \emph{Gaia} Science Operations Centre at ESAC for creating an excellent collaborative environment.}

\bibliographystyle{mn2e_AT}
\bibliography{AT_Mybib}

\appendix
\begin{appendix}
\section{On the choice of detection threshold}\label{sec:app1}
The three principal parameters that define the probability of astrometric orbit detection are the photocentric orbit size $\alpha$, the single-measurement uncertainty $\sigma$, and the number of measurements $N$. Several studies define a detection criterion {on the basis of a $\chi^2$ test in a simulation that includes the generation of synthetic observation} \citep{Sozzetti:2014qy,Casertano:2008th}, which in practice translates into $\alpha/\sigma>3$ for detected systems. 

We instead apply a detection criterion on the basis of $S/N = \alpha\, \sqrt{N_\mathrm{m}} / \sigma_\mathrm{m} >20$ for orbits with periods shorter than the measurement timespan. In all these considerations we neglect the potential effects of extreme eccentricities, which can reduce the astrometric signature by a factor $1-e^2$ in the worst case, and we assume a sufficient amount of degrees of freedom to solve the problem, i.e. about twice as many data points as free model parameters. 

\begin{figure}
\center
\includegraphics[width= \linewidth]{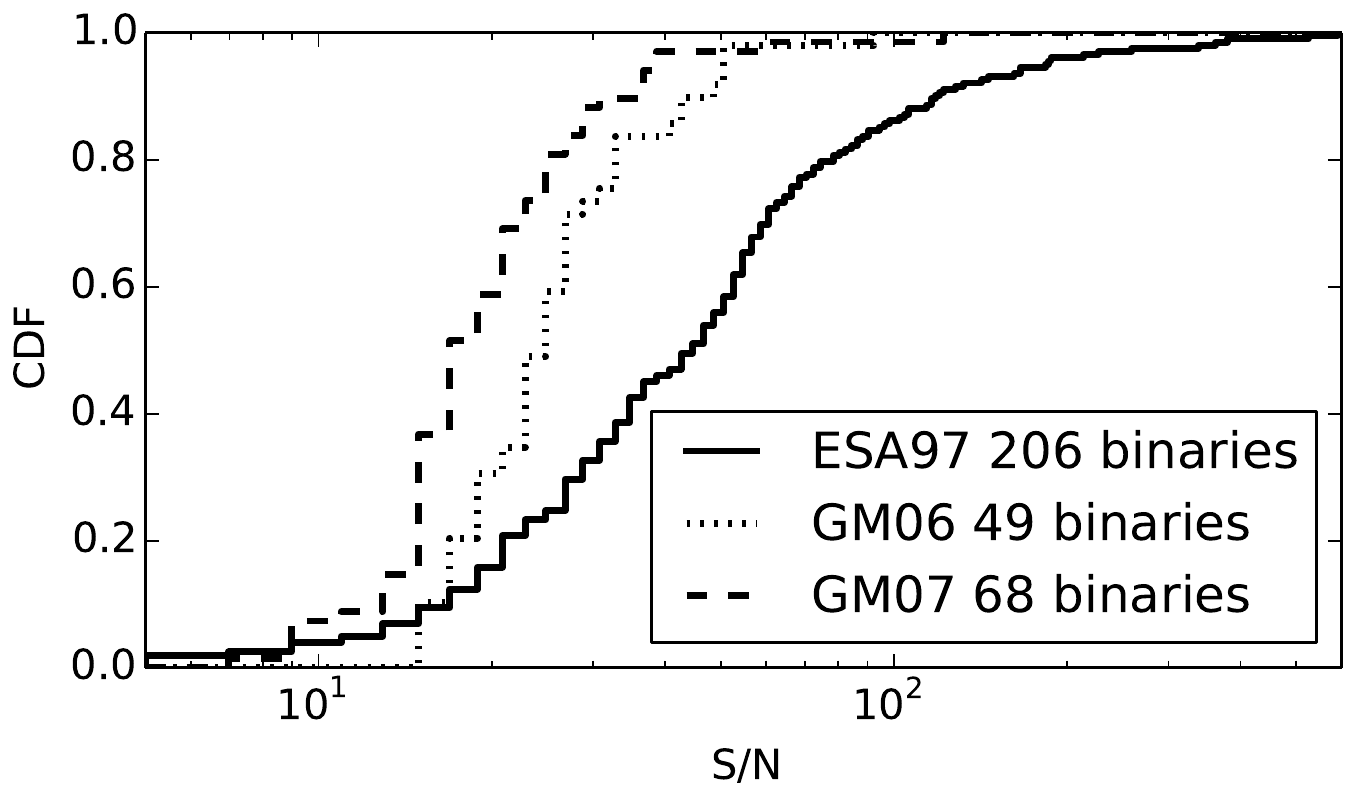}
\caption{Cumulative histogram of $S/N$ for \emph{Hipparcos} binary solutions published by \citet{ESA:1997vn} (ESA97), \citet{Goldin:2006fu} (GM06), and \citet{Goldin:2007ly} (GM07). The number of binaries is given in the legend.}
\label{fig:HIPSNR}
\end{figure}

Setting a detection threshold always involves a trade-off between detectability, false-alarm probability, and resulting parameter uncertainties. We tried to quantify the implications of our detection criterion $S/N>20$ by inspecting the binary detections with \emph{Hipparcos} data, which are analogous to exoplanet and binary detections with \emph{Gaia}. We used three catalogues, the original \emph{Hipparcos} double and multiple catalog \citep{ESA:1997vn} and the two binary catalogues of \cite{Goldin:2006fu} and \cite{Goldin:2007ly}. For every binary with period $<$1500 day, we computed the $S/N$ and the result is shown in Fig. \ref{fig:HIPSNR}. 

Only 11 \% of the \citet{ESA:1997vn} binaries were found with $S/N<20$. Using more sophisticated methods, \citet{Goldin:2006fu} and \citet{Goldin:2007ly} could increase this fraction to 18 \% and 44 \% for their respective samples of \emph{Hipparcos} stars with `stochastic` astrometric solutions.

\begin{figure}
\center
\includegraphics[width= \linewidth]{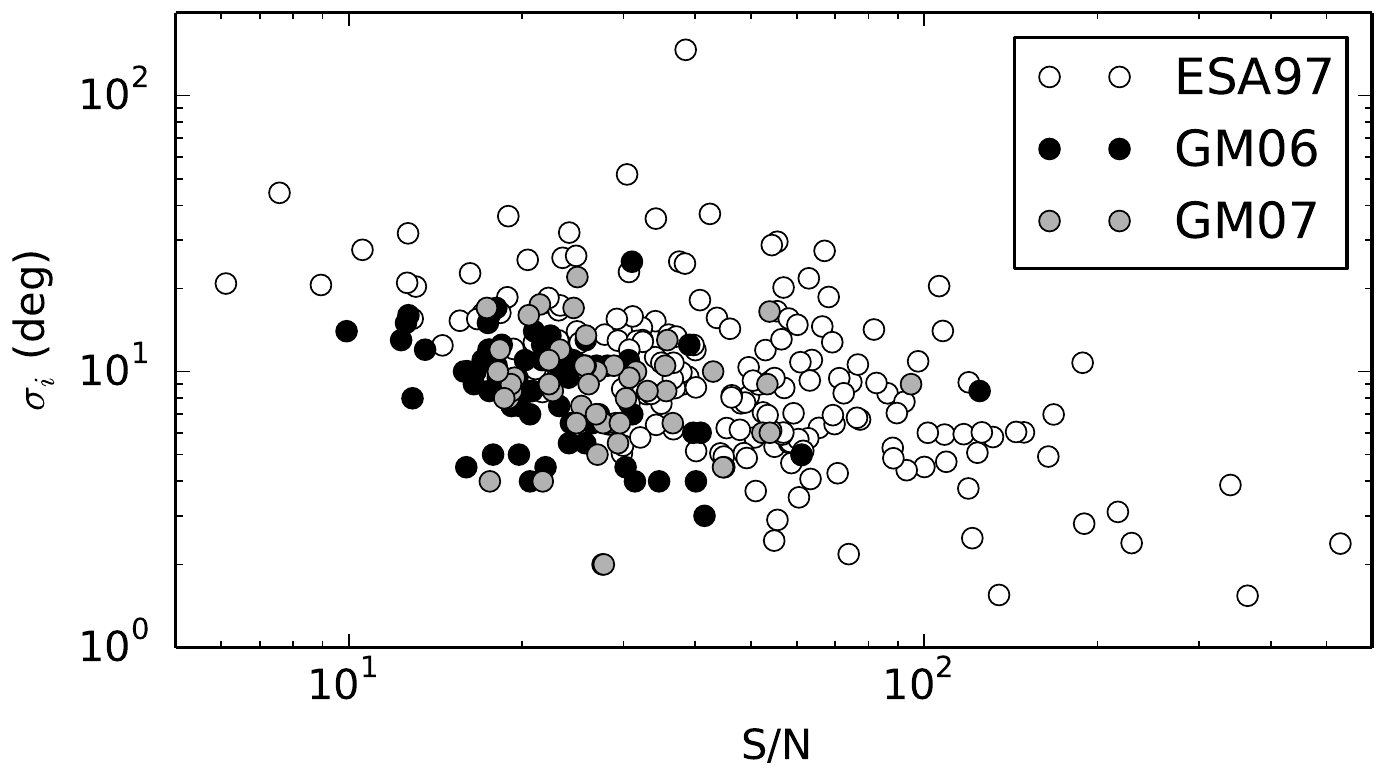}
\caption{Uncertainty in the orbit's inclination as a function of $S/N$. }
\label{fig:HIPinc}
\end{figure}

A figure of merit of particular interest for circumbinary planets is the uncertainty with which the orbital inclination can be determined. Figure \ref{fig:HIPinc} shows the inclination uncertainty $\sigma_i$ for the binary solutions in the literature. We see that for $S/N>20$, the inclination is typically determined to better than 10\degr. For $S/N$ larger than 100, the inclination uncertainty drops to a few degrees. 

We conclude that a threshold of $S/N>20$ leaves a safe margin in terms of false-alarm probability and provides us with an inclination uncertainty of $\lesssim10\degr$. {It also has the advantage of being easy to implement and fast to compute.}

\section{Additional figures and tables}

\begin{figure}
\center
\includegraphics[width= \linewidth]{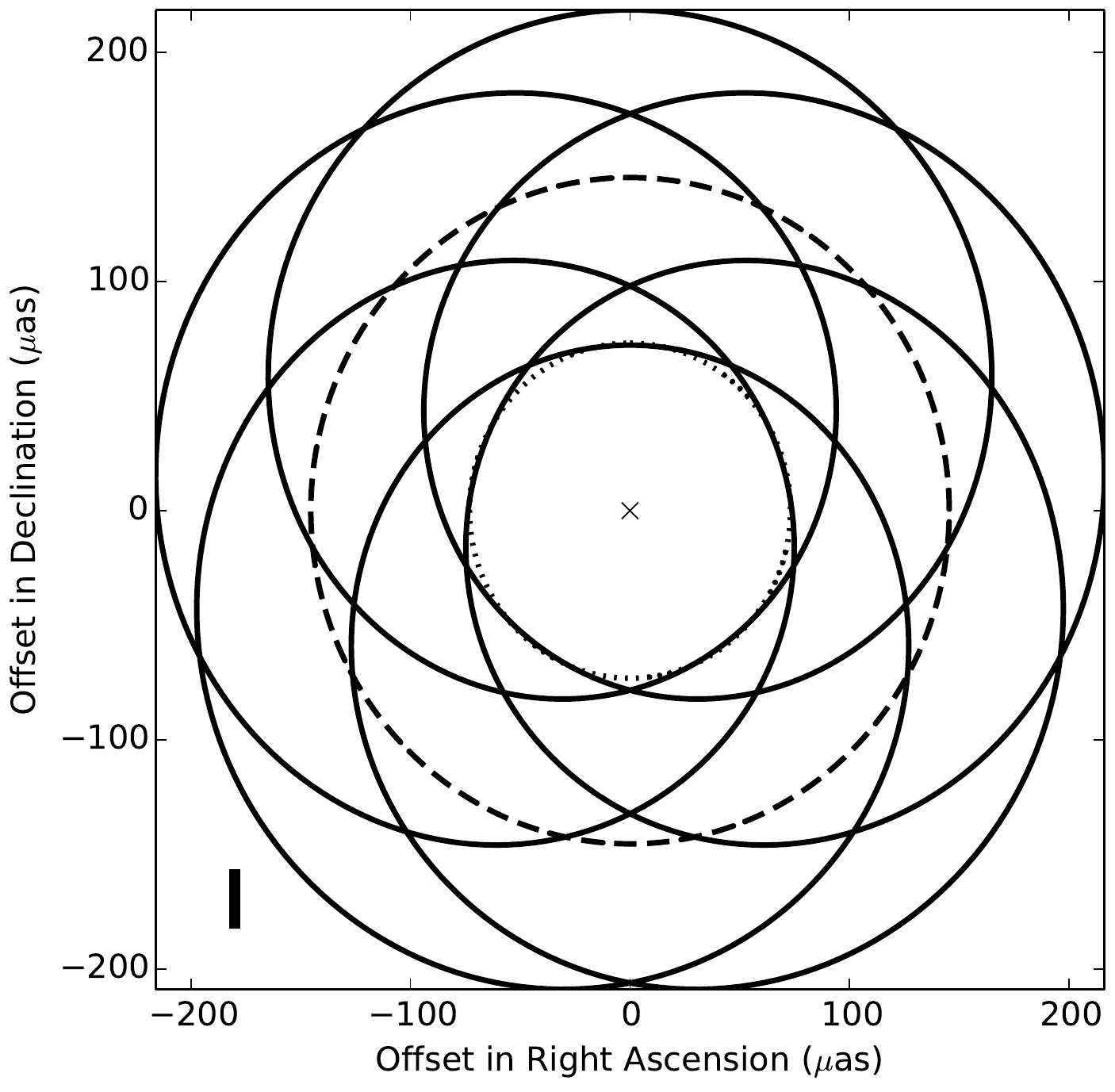}
\caption{{Example of the photocentre motion for a circumbinary planet detected with $S/N = 20$ taken from our simulation. For simplicity, we show a face-on and coplanar configuration, where parallax and proper motion have been removed. The $0.96+ 0.94 \,M_{\sun}$ binary with an orbital period of 40 days at 50 pc distance has a photocentric amplitude of $\alpha=145 \,\mu$as (dashed line). The planet with a mass of 7.9 $M_\mathrm{J}$ induces orbital motion with $73\,\mu$as amplitude and 240 days period (dotted line). The combined photocentre motion about the centre of mass (`x`) is shown as a solid curve. The bar in the lower-left corner indicates the single-measurement uncertainty of $31 \,\mu$as.}}
\label{fig:CBP_signature}
\end{figure}

\begin{figure}
\center
\includegraphics[width= \linewidth]{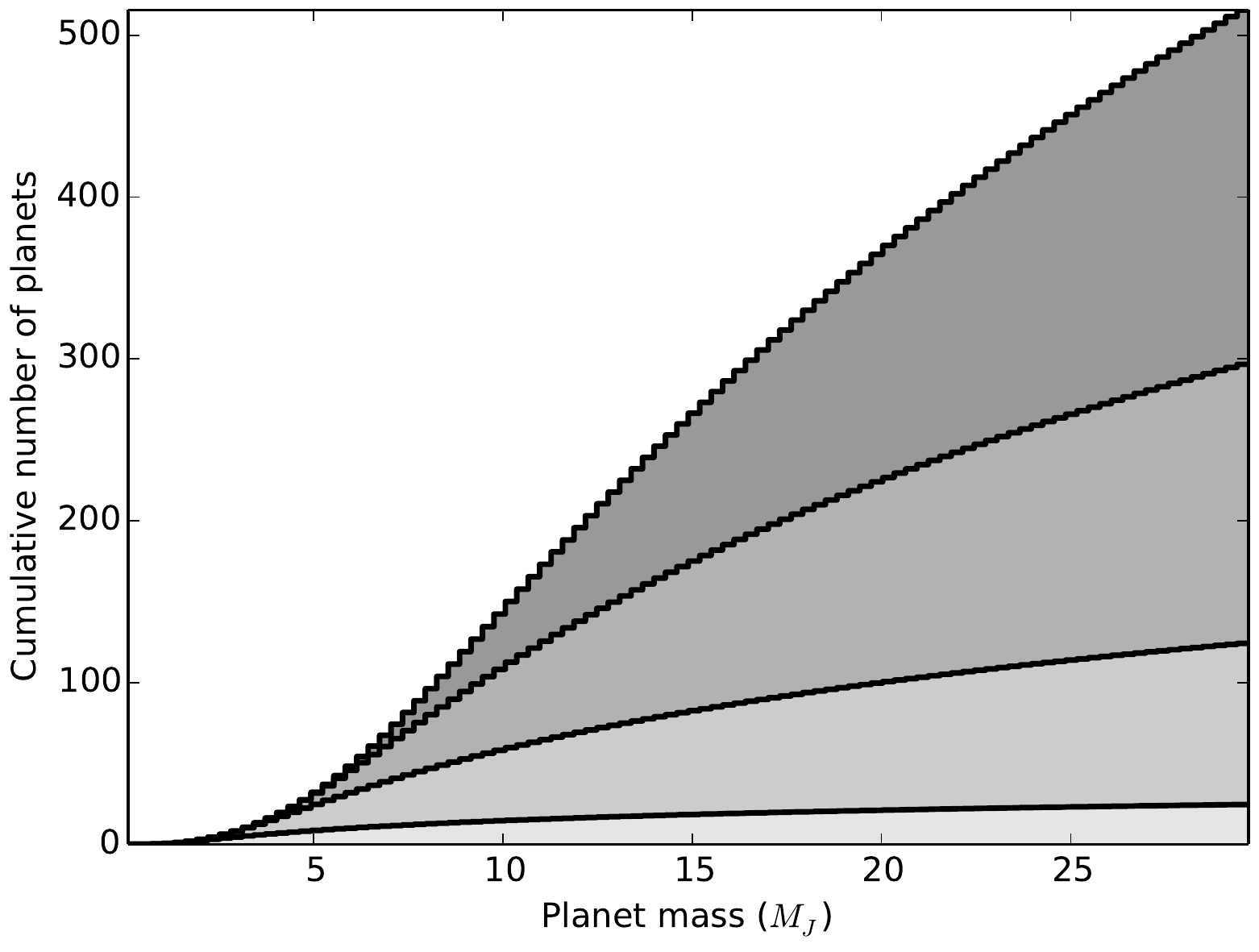}
\caption{Cumulative mass histogram of {detectable} circumbinary planets. The greyscale coding is like in Fig. \ref{fig:histomass}.}
\label{fig:histomasscum}
\end{figure}
\begin{figure}
\center
\includegraphics[width= \linewidth]{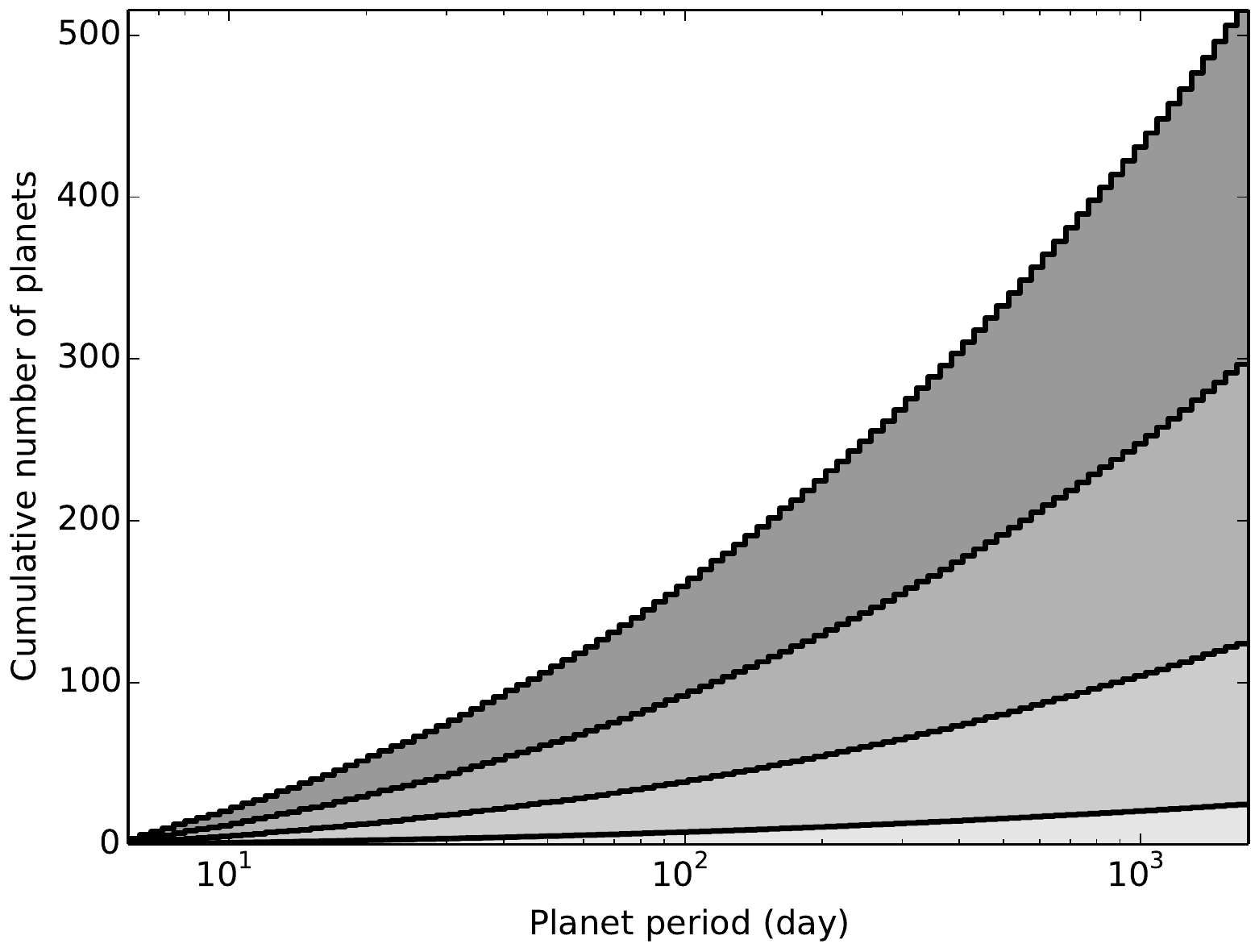}
\caption{Cumulative period histogram of {detectable} circumbinary planets. The greyscale coding is like in Fig. \ref{fig:histomass}.}
\label{fig:histoperiodcum}
\end{figure}

\begin{table*}
\begin{minipage}{126mm}
\caption{Circumbinary planet candidates {discovered by eclipse timing} that have a distance measurement or estimate. The reference codes are (1)~\citet{Deeg:2008aa}; (2) \citet{Beuermann:2010fk}; (3) \citet{Qian:2010aa}; (4) \citet{Beuermann:2012fk}; (5) \citet{Qian:2010ab}; (7) \citet{Qian:2012aa}.}
\label{tab:nnser}
\begin{tabular}{cccrrrrrrcc}

\hline
Name & $M_{\rm bin}$ & $V$ & $d$ & $P_{\rm p}$ & $M_{\rm p}$ & $\sigma_t$ & $a_{\rm 1,p}$ & $S/N$ & $r_{\rm orb}$ & Ref. \\
 & ($M_{\sun}$) & (mag) & (pc) & (days) & ($M_{\rm J}$) & ($\mu$as)& ($\mu$as) & & &\\
 \hline
CM Dra & 0.44 & 12.9 & 16 & 6757 & 1.5 & 43 & 1085 & 212.0 & 0.27 & 1 \\
NN Ser & 0.65 & 16.6 & 500 & 2776 & 1.7 & 231 & 17 & 0.6 & 0.66 & 2 \\
NN Ser & 0.65 & 16.6 & 500 & 5661 & 7.0 & 231 & 110 & 4.0 & 0.32 & 2 \\
DP Leo & 0.69 & 17.5 & 400 & 8693 & 6.3 & 356 & 158 & 3.7 & 0.21 & 3 \\
HW Vir & 0.63 & 10.9 & 180 & 4639 & 14.3 & 30 & 554 & 153.3 & 0.39 & 4 \\
HW Vir & 0.63 & 10.9 & 180 & 20089 & 65.0 & 30 & 6372 & 1763.6 & 0.09 & 4 \\
QS Vir & 1.21 & 14.4 & 48 & 2885 & 6.4 & 84 & 443 & 44.3 & 0.63 & 5 \\
NY Vir & 0.6 & 13.3 & 710 & 2885 & 2.3 & 51 & 17 & 2.8 & 0.63 & 7 \\
\hline
\end{tabular}
\end{minipage}
\end{table*}

\end{appendix}

\bsp

\label{lastpage}

\end{document}